\documentclass{article}
\usepackage{spconf,amsmath,graphicx}
\usepackage{braket}
\usepackage{comment}
\usepackage{amsfonts} 

\title{Hybrid quantum-classical graph neural networks for tumor classification in digital pathology}
\newcommand*{\affmark}[1][*]{\textsuperscript{#1}}

\name{Anupama Ray\affmark[1]  \qquad Dhiraj Madan\affmark[1] \qquad Srushti Patil\affmark[3] \qquad Maria Anna Rapsomaniki\affmark[2] \qquad Pushpak Pati\affmark[2]}
\address{\affmark[1] IBM Quantum IBM Research India, 
   \affmark[2]IBM Research Zurich \\
   \affmark[3]Indian Institute of Science Education and Research Tirupati, India}

\begin{document}
%
\maketitle
\begin{abstract}
Advances in classical machine learning and single-cell technologies have paved the way to understand interactions between disease cells and tumor microenvironments to accelerate therapeutic discovery. However, challenges in these machine learning methods and NP-hard problems in spatial Biology create an opportunity for quantum computing algorithms. We create a hybrid quantum-classical graph neural network (GNN) that combines GNN with a Variational Quantum Classifier (VQC) for classifying binary sub-tasks in breast cancer subtyping. We explore two variants of the same, the first with fixed pretrained GNN parameters and the second with end-to-end training of GNN+VQC. The results demonstrate that the hybrid quantum neural network (QNN) is at par with the state-of-the-art classical graph neural networks (GNN) in terms of weighted precision, recall and F1-score. We also show that by means of amplitude encoding, we can compress information in logarithmic number of qubits and attain better performance than using classical compression (which leads to information loss while keeping the number of qubits required constant in both regimes). Finally, we show that end-to-end training enables to improve over fixed GNN parameters and also slightly improves over vanilla GNN with same number of dimensions.

\end{abstract}
\section{Introduction}
\label{sec:intro}

Understanding how tumor cells self-organize and interact within the tumor microenvironment (TME) is a long standing question in cancer Biology, with the potential to lead to more informed patient stratification and precise treatment suggestions. From Hematoxylin \& Eosin (H\&E) staining to multiplexed imaging and spatial omics, a plethora of technologies are used to interrogate the spatial heterogeneity of tumors \cite{lewis2021spatial}. For example, H\&E histopathology images have long been used to train Convolutional Neural Networks (CNNs) in a patch-wise manner for a variety of tasks \cite{abduljabbar2020geospatial, niazi2019digital}. More recently, geometric deep learning and in particular Graph Neural Networks (GNNs) have found promising applications in histopathology \cite{jaume2021histocartography, pati2022hierarchical}. 
Indeed, a graph representation is a natural modeling choice for TME as it is a flexible data structure to comprehensively encode the tissue composition in terms of biologically meaningful entities, such as cells, tissues, and their interactions.
In a typical cell-graph representation, cells represent nodes, edges represent cell-to-cell interactions and cell-specific information can be included as node feature vectors. 
As a result, GNNs can elegantly integrate cellular information with tumor morphology, topology, and interactions among cells and/or tissue structures \cite{li2022graph}. Yet, the complexity of tumor graphs and the entangled cell neighborhoods lead to sub-optimal embedding spaces of GNNs, which in turn struggle with learning clinically meaningful patterns from the data. At the same time, searching for relatively small query subgraphs over large, complex graphs is $NP$-hard. Although GNNs are currently being used as state-of-art networks for learning such problems from images, two severe limitations of GNNs are over-smoothing \cite{Chen_Lin_Li_Li_Zhou_Sun_2020} and over-squashing \cite{topping2022understanding}. Over-smoothing refers to the indistinguishable representations of nodes in different classes and over-squashing refers to the inefficient message passing in a longer chain of nodes in a graph. These challenges in classical GNNs provide opportunities for quantum algorithms. The main impact expected from quantum is the possibility of extending the embedding space by mapping data to the exponentially large qubit Hilbert space, which can potentially help in capturing hidden spatio-temporal correlations at the cellular and tissue level. 

In this paper we create a hybrid classical-quantum network which combines a GNN with a Variational Quantum Classifier (VQC). We train this network with two approaches: (i) a serial approach, i.e., by first training the classical model and then the quantum model after the classical model has converged, and (ii) an end-to-end approach, by back-propagating loss from quantum neural network to all the layers of the classical neural network. In the first approach, we pretrain a classical graph neural network on the tissue graphs and then use the learnt representation from the GNN as input to a VQC. 
Since we are taking the output of the final layer of the classical GNN, we could map it with different dimensions via a linear layer. We performed ablation studies with 10-, 64-, 256-, 512- and 1024-dimensional learned GNN embeddings. 
For the 10-dimensional GNN output, wherein the learnt embedding has been compressed classically, we use second-order Pauli encoding (ZZ encoding), which needs as many qubits as the number of dimensions (thus 10 qubit circuits). For all other dimensional embeddings, we use amplitude encoding to be able to fit all the information in size logarithmic in embedding dimension (thus number of qubits needed is $\log(n)$ for $n$-dimensional output of GNN). A key observation of this paper is that although amplitude encoding compresses the number of qubits significantly, it does not lead to information loss, suggesting that the quantum model could be as close to state-of-art classical model. 
However, the quantum models with ZZ encoding are unable to learn much due to lossy compression via classical network.
In the second end-to-end approach, we experiment with 10-dimensional data with ZZ encoding. We observe that not only does end-to-end training of GNN+VQC significantly improve over serial, but it even slightly outperforms classical GNN with 10-dimensional final layer.
\section{Related Work and Background}
\subsection{Quantum Computing and Quantum Machine Learning}
Quantum Computing is a model of computation which enables one to perform efficient computation based on the laws of quantum mechanics. Here, the fundamental building blocks constitute qubits and gates.
A single qubit $\ket{\psi}$ can be mathematically expressed as a unit vector in a 2-dimensional Hilbert space as $\ket{\psi} = \alpha \ket{0}+\beta \ket{1}$, where $\vert\alpha\vert^2 +\vert \beta \vert^2=1$. Here $\ket{0}$ and $\ket{1}$ are the orthonormal basis states corresponding to classical bits 0 and 1. Similarly, an $n$-qubit state can be expressed as a unit vector in $2^n$ dimensional space $\ket{\psi} =\sum_{x \in \{0,1\}^n} \alpha_x \ket{x}$.
A measurement of an $n$-qubit state yields one of the classical bit strings $x$ with probability $\vert\alpha_x\vert^2$. 
A quantum circuit starts from an initial state $\ket{0^n}$ and performs a sequence of single and 2 qubit operations such as H, S, T, X, Y, Z, CNOT to yield a final state $\ket{\psi}$.
The above gate set also includes parameterized gates, such as $R_x(\theta),  R_y(\theta)$ and $R_z(\theta)$.
The produced final state can be measured to yield an output from the desired distribution corresponding to the problem (\cite{nielsen2001quantum}). 

Quantum circuits can be parameterized by learnable parameters and can also be trained to optimize a given objective function. In the context of machine learning, these are known as Variational Quantum Classifiers or VQC \cite{mitarai2018quantum, farhi2018classification}, which define the objective function based on the cross-entropy loss between the sampled distribution and ground truth data for classification.
Here the state is produced by first running a unitary parameterized by the input on initial state (feature map) followed by a unitary parameterized with trainable weights. Overall, we have the state $\ket{\psi(x, \theta)} = V_\theta U_{\phi(x)} \ket{0}$.
Some common feature maps include for example the Pauli feature map \cite{havlivcek2019supervised} and amplitude encoding \cite{schuld2021supervised}. 
The Pauli feature map maps an input $x$ to a quantum state $U_{\phi(x)} \ket{0^n}$, where
$U_{\phi(x)} = exp(i\sum_{S \in \mathcal{I}} \phi_S(x) \prod_{i \in S} P_i)$. Here, $\mathcal{I}$ in a collection of Pauli strings and $S$ runs over the set of indices corresponding to qubits where Paulis are applied.  
Here $\phi_S(x)  =
\left\{
	\begin{array}{ll}
		x_i  & S={i} \\
		 \prod_{j \in S} (\pi-x_j)  & \mbox{if } \vert S \vert>1
	\end{array}
\right\}$.

A special case of the same is given by the ZZ Feature map. Multiple repetitions of Pauli and ZZ Feature maps can be stacked as well. Another common feature map is amplitude encoding, which encodes a vector $x \in \mathbb{R}^n$ as $\sum_i \frac{x_i}{\|x\|} \ket{i}$. This takes $\log(n)$ qubits  whereas ZZ encoding requires $n$ qubits. One can measure the state to obtain samples from the model distribution by measuring an observable $O$ on the state $p(y \vert x ; \theta) =  \braket{\psi(x, \theta)| O|\psi(x, \theta)}$.
One can take the observable to be $ZZ..ZZ$, which corresponds to measuring parity $\in \{+1,-1\}$. 
The cost function can be optimized using classical routines, e.g., COBYLA, SPSA, Adam, NFT.
\subsection{Classical Neural Networks for Spatial Tissue Modeling}
HACT-NET~\cite{pati2022hierarchical} is a state-of-the-art Graph Neural Network model for the hierachical analysis of digital pathology tissue images. 
Typically the tissue images are of large dimensions, e.g., 5000 $\times$ 5000 pixels at 40$\times$ magnification (0.46 $\mu$m/pixel). To process such images by a CNN while utilizing the complete TME context is infeasible due to the high computational overload. Therefore, a graph representation is  useful to encode the necessary TME information in terms of a thousands of nodes and edges, and is much lighter than a pixel-based image representation.
Building on this concept, HACT-NET constructs a hierarchical graph representation of a tissue by incorporating a low-level cell-graph, a high-level tissue-graph, and a cell-to-tissue hierarchy to comprehensively represent the tissue composition. 
Afterwards, the hierachical GNN backbone of HACT-NET processes the graph representation in a two-step manner to produce a cell- and tissue-aware feature embedding. A Multi-Layer Perceptron (MLP) operates on this embedding to perform downstream tissue subtyping.
In this work, we pre-train the HACT-NET model for various downstream tissue classification tasks and use the pre-trained model to extract tissue embeddings for subsequently training our VQC.
\section{Methodology}
\label{sec:pagestyle}
\begin{figure}
    \centering
    \includegraphics[scale = 0.47]{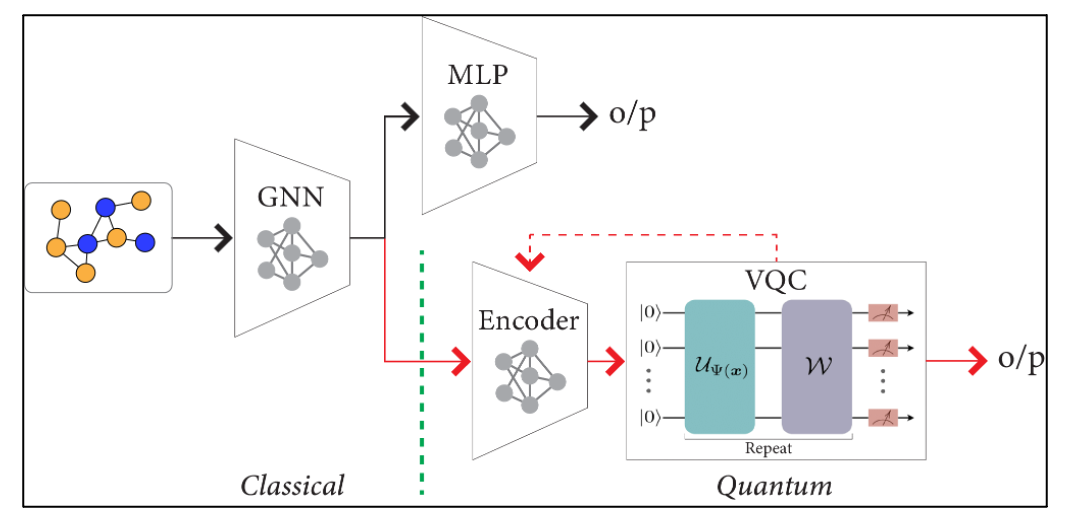}
    \caption{Implementation of hybrid GNN-VQC model}
    \label{vqc}
\end{figure}
In our approach, we define a hybrid classical-quantum graph neural network, an overview of which is shown in Figure \ref{vqc}. 

Specifically, we use a HACT-NET~\cite{pati2022hierarchical} to produce embeddings as $Embed(x; \theta_G)= GNN(x; \theta_G) \in \mathbb{R}^d$ corresponding to the input image $x$. These embeddings are then passed as input to a VQC which applies a feature map followed by an ansatz $V_{\theta_Q}$ and produces samples from the distribution 
\begin{equation}
p(y \vert x ; \theta_G, \theta_Q) =  \braket{\psi(x; \theta_G, \theta_Q)| ZZ...ZZ|\psi(x; \theta_G, \theta_Q)}
\label{eq:1}
\end{equation}
 \begin{equation}
\text{ where, } \ket{\psi(x; \theta_G , \theta_Q)}= V_{\theta_Q} U_{\phi(Embed(x; \theta_G))} \ket{0}.
 \label{eq:2}
 \end{equation}

 Here $\theta_G$ and $\theta_Q$ refer to GNN and VQC parameters respectively.

We follow two approaches for training our Hybrid Network: (i) with a pretrained GNN (having frozen weights), and (ii) with trainable GNN parameters. In the first approach, we first pretrain HACT-NET  with a classical MLP layer and then use the learnt representation of the final layer as input to the quantum network as defined in Equations \ref{eq:1} and \ref{eq:2}.
Here, the parameters $\theta_G$ are kept fixed after the initial pre-training stage. 
In the second approach, both sets of parameters are updated together. We discuss the details of second approach in section \ref{sec:endtoend} and focus on the first approach in this section.


When trained separately, the HACT-NET performed best at 64-dimensional output of GNN passed to the MLP before the final output. However, it is very difficult to get reliable results using 64 qubits in the current available quantum devices, which both have few qubits and the qubits are noisy. Thus, we experimented with a range of dimensions and different encoding schemes to use different number of qubits on the same data. We experimented with 10-dimensional output of GNN wherein we used ZZ encoding with 2 layers of repetition\cite{schuld2021supervised}. Here, number of qubits used equals the dimension.  
We also trained with higher embedding dimensions from the HACT-NET, such as 64, 256, 512 and 1024 with amplitude encoding. 

Thus, we were able to encode a 64-dimensional input in 6 qubits. With this encoding, we were able to reach the state-of-art classification F1-score that the GNN achieved. Since these classical neural networks have large number of parameters, they are known to overfit at higher dimensions in presence of less data. Since data shortage in a known limitation in most tissue imaging datasets, a key research question here is \textbf{can quantum models outperform classical models at higher dimensions where classical models tend to overfit}. In order to study this, we experimented with 256-, 512- and 1024-dimensional learnt representations of the GNN which were both passed to the classical MLP as well as the quantum classifier to study the effects of high dimensions. Using amplitude encoding we were able to encode these in 8, 9, and 10 qubits, respectively.

\subsection {Dataset}
For this work, we experimented on 3 binary classification tasks under the breast cancer sub-typing problem on the BReAst Cancer Subtyping (BRACS) dataset~\cite{pati2022hierarchical}. In BRACS, each image is of the order of 2048×1536 pixels and there are $\approx$2200 such images. We randomly split them into 1200 for training, 500 for validation and 500 for testing. 
\subsection{Training details}
In this subsection we explain the details of the VQC scheme. We apply parity-postprocessing after measurement (corresponding to measuring the observable $ZZ...Z$ on the parameterized state produced) to get the desired output and pass through a cost function. We update the parameters of the ansatz to minimize the overall cost function, much like training weights of a neural network. In the current implementation, the measurement results were interpreted based on the parity of the measurement outputs, where even parity is considered as label +1 and odd parity as -1. After obtaining labels from parity post-processing, the classical optimizer calculates the cost function, and optimizes the parameters of ansatz until the classical optimization iterations complete or until the cost function converges. For inference, we use multiple shots and the most probable label is selected as the final label for each test data. We trained our models with Constrained Optimisation By Linear Approximation (COBYLA) \cite{powell2007view} and Nakanishi-Fujii-Todo (NFT) \cite{Nakanishi_2020} optimizers and discuss the best results across both optimizers. 
The maximum number of epochs was set to 100 with early stopping. 
All our experiments with different data sizes are run on a noiseless state vector simulator provided by IBM Quantum. 
\subsection{End-to-end training}
\label{sec:endtoend}
For the end-to-end training, we train the parameters of GNN namely, $\theta_G$ and VQC parameters namely $\theta_Q$ together using Qiskit's TorchConnector class. We trained the above with 10-dimensional GNN embeddings using ZZ encoding for VQC. Since the classical neural networks trains using gradient based backpropagation, we use Adam optimizer for training both the networks with a learning rate of $10^{-3}$ for VQC parameters and $10^{-6}$ for GNN parameters. We found it useful to optimize the VQC parameters less frequently (once every 10 epochs) than the GNN parameters.


\section{Results}
\label{sec:majhead}

\begin{figure}
    \centering
    \includegraphics[scale = 0.4]{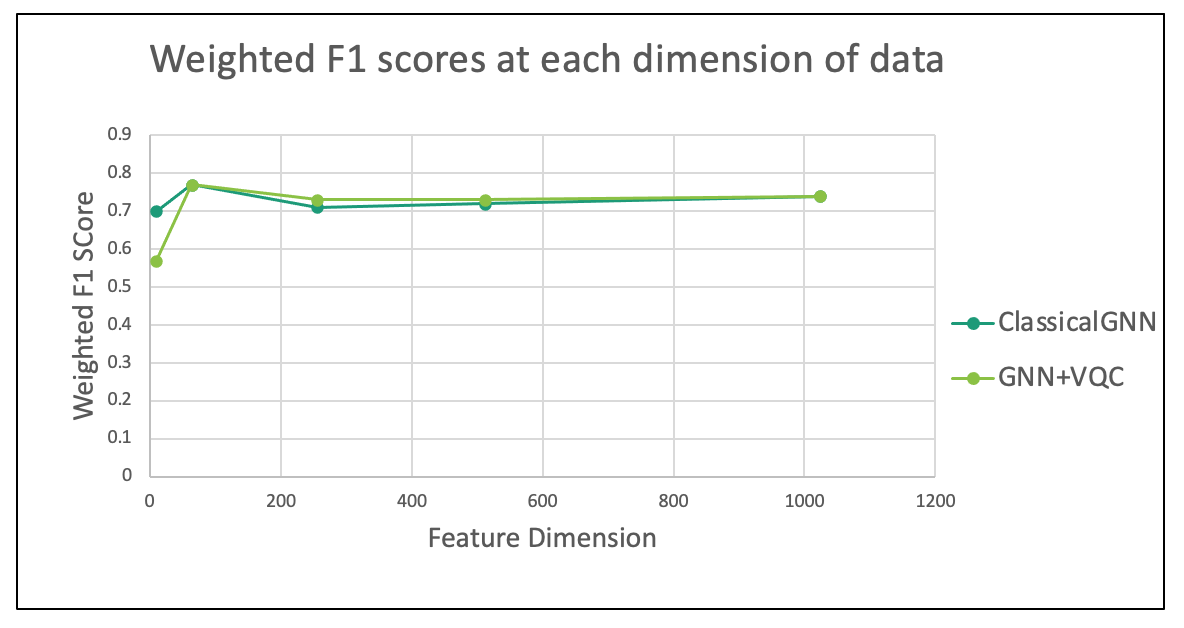}
    \caption{Graph showing performance (weighted F1-score) of classical GNN and hybrid quantum-classical model on different feature dimensions}
    \label{graph}
\end{figure}

In this section we present the results obtained by the hybrid quantum-classical model using different feature dimensions and embedding methods and its comparison to state-of-art classical GNN. We also present detailed ablation studies to understand the impact of training data sizes in training both classical GNN and the proposed hybrid model. Figure \ref{graph} shows the performance of classical GNN (in dark green) and hybrid quantum model (in light green) on different dimensional learnt embeddings. While at lower dimensions (10 and 64) classical GNN is able to learn better than the quantum model, the quantum model is at par with classical in higher dimensions of 256, 512 and 1024. 

\begin{figure}
    \centering
    \includegraphics[scale = 0.4]{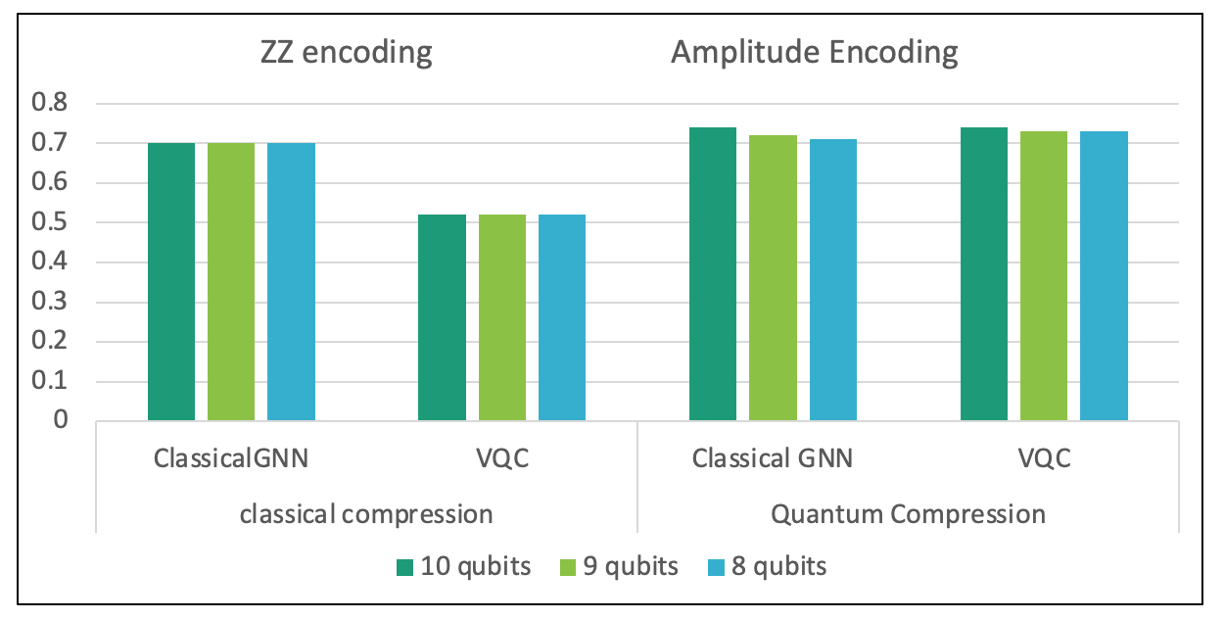}
    \caption{Classical compression vs Quantum compression}
    \label{comp}
\end{figure}

We further experiment in this direction to understand the difficulties in learning. While keeping the number of qubits constant, we change the encoding schemes to understand the impact of data compression. Figure \ref{comp} shows impact of classical vs quantum compression by means of changing different feature dimensions and accordingly choosing encoding schemes to represent them in quantum states. When we compress the data classically by reducing the number of output neurons to 10, 9 and 8 dimensions, we observe that although we use 10, 9 and 8 qubits respectively via ZZ encoding, the quantum model is unable to learn and struggles at a weighted F1-score of ~50\%. This is primarily due to the information loss in the neural network that happens during the classical compression. When the data is not classically compressed and we pass a feature representation of dimension 1024, or 512 or 256 represented by same 10, 9 and 8 qubits, then the quantum model is at par with the state-of-art classical model. Here we use amplitude encoding which encodes $n$ classical bits in $\log(n)$ qubits but does not lose any information, enabling the quantum model to learn better from the high dimensional data.  

Since classical deep learning networks are known to under-perform in low data scenarios, we wanted to study the impact of training data for both classical and quantum models. We perform a series of experiments wherein we use 0.1, 0.25, 0.5 and then full data for training both models. As expected, in both scenarios and across all dimensions, we observe that training with full data leads to the best results on the held-out test data, and the performance comparison trend is identical to the best model with full training. 

We also show the test results (weighted precision, weighted recall and weighted F1-score) on end-to-end training, in comparison with classical GNN as well as separately trained GNN+VQC in Table \ref{tab: mean and sd}. We show that end-to-end training significantly improves over separate training of VQC and GNN, and even slightly outperforms classical GNN. 

\begin{table}[h]
    \caption{Table comparing end-to-end trainable networks vs classicalGNN and classicalGNN+VQC trained separately. All experiments on 10-dimensional ZZ encoding using 10 qubits on the simulator.}
    \vspace{2mm}
    \centering
    \label{tab: mean and sd}
    \begin{tabular}{c| c c c }
    \hline
    \hline
    Model & w-precision & w-Recall & w-F1score\\
    \hline
    cGNN & 0.71	& 0.69	& 0.7 \\
    cGNN+VQC & 0.58	& 0.57	& 0.57\\
    end-to-end GNN+VQC & 0.72 & 0.71 & 0.72\\
    \hline
    \hline
    \end{tabular}  
\end{table}

\section{Discussions and Future Work}
Overall, in this work we present two ways to train hybrid quantum-classical neural networks. We show that end-to-end training is significantly better than serially training such models and demonstrate results on a real-world breast-cancer sub-typing task. In detailed ablation studies we observe that quantum compression can be significantly better to qubit requirements without information loss unlike lossy classical compression. Future directions could be to explore how other such classical networks can be combined with quantum circuits to enhance their trainability and improve generalization. 




\bibliographystyle{plain}
\bibliography{refs}

\end{document}